\documentclass[prd,12pt]{revtex4}

\usepackage{epsfig}
\usepackage{graphicx}
\usepackage{dcolumn}
\usepackage{amsmath}
\usepackage{latexsym}
\usepackage{color}
\usepackage{subfig}

\usepackage{footnote}
\usepackage{hyperref}
\hypersetup{
    colorlinks=true,
    linkcolor=blue,
    citecolor=blue,
    filecolor=blue,      
    urlcolor=blue,
    hyperfootnotes=true
}
 
\begin{document}

%%%%%%%%%%%%%%%%%%%%
\title{Analytic investigation of rotating holographic superconductors}  
%\date{\today}

\author{Ankur Srivastav\footnote{
\href{mailto:ankursrivastav@bose.res.in}{ankursrivastav@bose.res.in}}, Sunandan Gangopadhyay\footnote{\href{mailto:sunandan.gangopadhyay@gmail.com}{sunandan.gangopadhyay@gmail.com}\\       \href{mailto:sunandan.gangopadhyay@bose.res.in}{sunandan.gangopadhyay@bose.res.in}}}
\affiliation{Department of Theoretical Sciences, S. N. Bose National Centre for Basic Sciences,\\ Block-JD, Sector-III, Salt Lake City,\\ Kolkata 700106, India}

%%%%%%%%%%%%%%%%
\begin{abstract}
\noindent In this paper we have investigated, in the probe limit, $s$-wave holographic superconductors in rotating $AdS_{3+1}$ spacetime using the matching method as well as the St{\"u}rm-Liouville eigenvalue approach. We have calculated the critical temperature using the matching technique in such a setting and our results are in agreement with previously reported results obtained using the St{\"u}rm-Liouville approach. We have then obtained the condensation operators using both analytical methods. The results obtained by both these techniques share the same features as found numerically. We observe that the rotation parameter of the black hole affects the critical temperature and the condensation operator in a non-trivial way.
%%%%%%%%%%%%%%%%

\end{abstract}

\maketitle

\noindent Keywors: Rotating holographic superconductors,  gauge/gravity duality.

\section{Introduction}
\noindent High-$T_{c}$ superconductors were discovered by George Bednorz and K. Alex M{\"u}ller in 1986~\cite{bm}. These are prototype of the so called strongly correlated systems in condensed matter physics. Such systems generically fall under strongly coupled field theories. However, due to strong coupling these systems are hard to tame using traditional field theoretic approaches. In the last decade, the AdS/CFT correspondence has emerged as a powerful tool to study such systems. The AdS/CFT correspondence originated from string theory and was discovered by Maldacena in 1997 \cite{jm}. The conjecture implies that a gravity theory in a ($d+1$)-dimensional anti-de Sitter (AdS) spacetime is an exact dual to a conformal field theory (CFT) living on the $d$-dimensional boundary of the bulk spacetime. 

  The importance of the gauge/gravity correspondence in strongly coupled systems was realized in 2008 when Gubser showed that for an asymptotic AdS black hole, near to its horizon, $U(1)$ gauge symmetry spontaneously breaks giving rise to the phenomenon of superconductivity in the vicinity of the black hole horizon \cite{SSG}. Immediately using this result Hartnoll, Herzog, and Horowitz used the AdS/CFT correspondence to study holographic superconductors which mimics the properties of high-$T_{c}$ superconductors~\cite{hhh,hhh1}. Since then various aspects of such holographic superconductors have been explored in various black hole space-time settings \cite{CPH, SAH, GR1, GR2, SG1, GG1}. Most of the holographic superconductors considered so far are constructed using non-rotating black hole spacetime. However, even before the discovery of AdS/CFT correspondence, rotating black holes were known to show Meissner like effect \cite{wald, KM}. This motivates us to investigate about the role of rotating black holes in holographic superconductor models. Such a study was first initiated in \cite{js} where the study of spontaneous symmetry breaking in $3+1$-dimensional rotating, charged AdS black hole was carried out. It was observed that the superconducting order gets destroyed for a particular value of the rotation just below the critical temperature. Recently, a holographic superconductor model in rotating black hole spacetime was studied in \cite{la}. Here, the St{\"u}rm-Liouville (SL) eigenvalue approach was used to obtain the critical temperature analytically. However, the condensation operator was studied numerically. Our aim in this paper is to provide an analytical approach to rotating holographic superconductors and obtain the critical temperature as well as the condensation operator. We shall first use the matching technique to obtain the critical temperature for both possible values of the conformal dimension. This technique was introduced in \cite{GKS} and involves the matching of the solutions to the matter and gauge field equations near the black hole horizon and the AdS boundary. We then compare our results with those obtained from the SL technique in \cite{la}. Using the matching approach, we then obtain the values of the condensation operator. Further, we also apply the SL eigenvalue approach to obtain the condensation values and compare the results with those obtained by us using the matching technique and also with the numerical results in the literature. All the calculations in this paper have been done in the probe limit where we do not consider the backreaction of the matter sector into the metric in our model.      
 
  The paper is organized in the following way. In section \hyperlink{sec2}{II}, we set up a simple model for holographic superconductors in an uncharged rotating black hole background in $AdS_{3+1}$ spacetime. In section \hyperlink{sec3}{III}, we calculate the critical temperature and the condensation operator values using matching method, where we match the solutions to the field equations obtained at the boundaries at some appropriate point between the boundaries. Then in section \hyperlink{sec4}{IV}, we go on to calculate the critical temperature as well as the condensation operator values using the SL eigenvalue analysis. In the last section (\hyperlink{sec5}{V}) of this paper, we summarized our findings. 
\section{Setting Up of the Holographic Superconductor}
\noindent \hypertarget{sec2}{We} start by writing down the metric of an uncharged rotating black hole in $AdS_{3+1}$ spacetime \cite{as,la}
\begin{eqnarray}
 ds^{2}=-f(r)(\Xi dt-ad\varphi)^{2}+r^{2}(adt-\Xi d\varphi)^{2}+\dfrac{dr^{2}}{f(r)}+r^{2}d\theta^{2}
\label{metric}
\end{eqnarray}
where $$f(r)=\bigg(r^{2}-\dfrac{r_0^{3}}{r}\bigg) ~ , ~   \Xi=(1+a^{2})$$ with $r_{0}$  being the event horizon of the black hole and $a$ being the rotation parameter of the black hole. For convenience, we take unit AdS radius and the cosmological constant $\Lambda=-3$. The Hawking temperature associated with the above black hole geometry is given by \cite{la}
\begin{eqnarray}
T=\dfrac{3r_{0}}{4\pi\varrho}
\label{Hawking_temperature}
\end{eqnarray}
\begin{eqnarray}
\nonumber
~~~~~~~\varrho = \dfrac{\Xi}{\Xi^{2}-a^{2}}~. 
\end{eqnarray}
We now write down the simplest model for holographic superconductor with the Lagrangian density consisting of a Maxwell field and a complex scalar field minimally coupled to $A_{\mu}$. This reads 
\begin{eqnarray}
~~~~~~~~~ \mathcal{L}=-\dfrac{1}{4}F^{\mu\nu}F_{\mu\nu}-|D\Psi|^{2}-m^{2}|\Psi|^{2}
\label{Lagrangian_density}
\end{eqnarray}
$$F_{\mu\nu} \equiv \partial_{[\mu}A_{\nu]} ~ , ~ D \equiv (\partial-iA)~.$$
Since $g_{\mu\nu}$ depends only on $r$, we set $\Psi=\psi(r)$. We also make the ansatz $A_{\mu}=\delta^{t}_{\mu}\Phi(r)+\delta^{\varphi}_{\mu}\Omega(r)$ because of the presence of the $g_{t\phi}$ term in metric (\ref{metric}). Now varying the Lagrangian density $\mathcal{L}$, we get the following equations for the matter field $\psi(r)$ and the gauge fields $\Phi(r)$ and $\Omega(r)$
\begin{eqnarray}
    \psi^{\prime\prime}+\bigg(\dfrac{f^{\prime}}{f}+\dfrac{2}{r}\bigg)\psi^{\prime}+\bigg[\bigg(\dfrac{(1+a^{2}) \Phi+a \Omega}{f(1+a^{2}+a^{4})}\bigg)^{2}-\dfrac{1}{f}\bigg(\dfrac{a \Phi+(1+a^{2}) \Omega}{r(1+a^{2}+a^{4})}\bigg)^{2}-\dfrac{m^{2}}{f}\bigg]\psi=0
\label{Psi_field_eq_1}
\end{eqnarray}
\begin{eqnarray}
    \Phi^{\prime\prime}+\bigg(\dfrac{2(1+a^{2})^{2}}{r(1+a^{2}+a^{4})}-\dfrac{a^{2}f^{\prime}}{f(1+a^{2}+a^{4})}\bigg)\Phi^{\prime}+\bigg(\dfrac{2a(1+a^{2})}{r(1+a^{2}+a^{4})}-\dfrac{a(1+a^{2})f^{\prime}}{f(1+a^{2}+a^{4})}\bigg)\Omega^{\prime}-\dfrac{2\psi^{2}}{f}\Phi=0 \nonumber \\
\label{Phi_field_eq_1}
\end{eqnarray}
\begin{eqnarray}
    \Omega^{\prime\prime}-\bigg(\dfrac{2a^{2}}{r(1+a^{2}+a^{4})}-\dfrac{(1+a^{2})^{2}f^{\prime}}{f(1+a^{2}+a^{4})}\bigg)\Omega^{\prime}-\bigg(\dfrac{2a(1+a^{2})}{r(1+a^{2}+a^{4})}-\dfrac{a(1+a^{2})f^{\prime}}{f(1+a^{2}+a^{4})}\bigg)\Phi^{\prime}-\dfrac{2\psi^{2}}{f}\Omega=0 \nonumber \\
\label{Omega_field_eq_1}
\end{eqnarray}
where $^{\prime}$ denotes derivative with respect to $r$.\\
 We now make the change of coordinates  $z=\dfrac{r_{0}}{r}$ due to which the  field eq.(s)(\ref{Psi_field_eq_1}, \ref{Phi_field_eq_1}, \ref{Omega_field_eq_1}) take the form 
\begin{eqnarray}
z\psi^{\prime\prime}-\bigg(\dfrac{2+z^{3}}{1-z^{3}}\bigg)\psi^{\prime}+\bigg(\dfrac{z((1+a^{2})\Phi + a\Omega)^{2}}{(1+a^{2}+a^{4})^{2}r_{0}^{2}(1-z^{3})^{2}} + \dfrac{z(a\Phi + (1+a^{2})\Omega)^{2}}{(1+a^{2}+a^{4})^{2}r_{0}^{2}(z^{3}-1)}-\dfrac{m^{2}}{z(1-z^{3})}\bigg)\psi=0 \nonumber \\
\label{Psi_field_eq_2}
\end{eqnarray}
\begin{eqnarray}
\Phi^{\prime\prime}+\dfrac{3a^{2}z^{2}}{(1+a^{2}+a^{4})(1-z^{3})}\Phi^{\prime}-\dfrac{2\psi^{2}}{z^{2}(1-z^{3})}\Phi + \dfrac{3a(1+a^2)z^{2}}{(1+a^{2}+a^{4})(1-z^{3})}\Omega^{\prime}=0 
\label{Phi_field_eq_2}
\end{eqnarray}
\begin{eqnarray}
\Omega^{\prime\prime}-\dfrac{3(1+a^{2})^{2}z^{2}}{(1+a^{2}+a^{4})(1-z^{3})}\Omega^{\prime}-\dfrac{2\psi^{2}}{z^{2}(1-z^{3})}\Omega - \dfrac{3a(1+a^{2})z^{2}}{(1+a^{2}+a^{4})(1-z^{3})}\Phi^{\prime}=0
\label{Omega_field_eq_2}
\end{eqnarray}
where $^\prime$ denotes derivative with respect to $z$.
\section{Critical temperature and condensation values using matching method}
\noindent \hypertarget{sec3}{In} this section, we shall apply the  so called matching method~\cite{sg} to find the critical temperature and the condensation operator values. In this approach one writes down the approximate solutions of the field equations for $\psi, \Phi$ and $\Omega$ near the horizon and around the AdS boundary and then match them at some point to determine the unknown coefficients in the solutions. Before we proceed, we note that the finiteness of $A_{\mu}$ at the horizon gives the boundary conditions $\Phi(1)=0$ and $\Omega(1)=0$. We carry out our analysis for $m^{2}=-2$ which is within the Breitenlohner--Freedman (BF) mass bound~\cite{BF1, BF2}. 
\subsection{Solution near the event horizon ($z = 1$)}
\noindent We first Taylor expand the fields $\psi(z), \Phi(z)$ and $\Omega(z)$ as
\begin{eqnarray}
~~~~~~~~~\psi(z) = \psi(1)-\psi^{\prime}(1)(1-z)+\dfrac{1}{2}\psi^{\prime\prime}(1)(1-z)^{2}+. . .
\label{Psi_naer_z_1}
\end{eqnarray}
\begin{eqnarray}
~~~~~~~~~\Phi(z) = \Phi(1)-\Phi^{\prime}(1)(1-z)+\dfrac{1}{2}\Phi^{\prime\prime}(1)(1-z)^{2}+ . . . 
\label{Phi_near_z_1}
\end{eqnarray}
$$ \approx  -\Phi^{\prime}(1)(1-z)+\dfrac{1}{2}\Phi^{\prime\prime}(1)(1-z)^{2}          $$
\begin{eqnarray}
~~~~~~~~~\Omega(z) = \Omega(1)-\Omega^{\prime}(1)(1-z)+\dfrac{1}{2}\Omega^{\prime\prime}(1)(1-z)^{2}+ . . . 
\label{Omega_near_z_1}
\end{eqnarray}
$$\approx -\Omega^{\prime}(1)(1-z)+\dfrac{1}{2}\Omega^{\prime\prime}(1)(1-z)^{2}~. $$
In writing down these expressions, we have used the boundary conditions $\Phi(1)=0$ and $\Omega(1)=0$. Now from eq.(\ref{Psi_field_eq_2}), we obtain in the $z \to 1$ limit  
\begin{eqnarray} 
\psi^{\prime}(1)=\dfrac{2}{3}\psi(1)~.
\label{psi_prime_z_1}
\end{eqnarray}
Further eq.(\ref{Phi_field_eq_2}) in the limit $z \to 1$ leads to 
\begin{eqnarray} 
a\Phi^{\prime}(1)+(1+a^{2})\Omega^{\prime}(1)=0~.
\label{Phi_Omega_prime_z_1}
\end{eqnarray}
From eq.(s)(\ref{Psi_field_eq_2}, \ref{Phi_field_eq_2}, \ref{Omega_field_eq_2}), in the limit $z \to 1$, we also obtain 
\begin{eqnarray}
~~~~~~~~~~~~~~~~~~~~~\psi^{\prime\prime}(1) = - \bigg[\dfrac{4}{9}+\dfrac{((1+a^{2})\Phi^{\prime}(1) +a \Omega^{\prime}(1))^{2}}{18r_{0}^{2}(1+a^{2}+a^{4})^{2}} \bigg]\psi(1)
\label{1_psi_prime_prime_z_1}
\end{eqnarray}
\begin{eqnarray}
\Phi^{\prime\prime}(1)=-\dfrac{2}{3}\psi^{2}(1)\Phi^{\prime}(1)f(a)
\label{1_Phi_prime_prime_z_1}
\end{eqnarray}
\begin{eqnarray}
\Omega^{\prime\prime}(1)=-\dfrac{2}{3}\psi^{2}(1)\Omega^{\prime}(1)f(a)
\label{1_Omega_prime_prime_z_1}
\end{eqnarray}
where $$f(a) = \dfrac{(1+a^{2}+a^{4})^{2}}{(1+2a^{2}+3a^{4}+2a^{6}+a^{8})}~.$$
The Taylor series expansions for the fields $\psi(z), \Phi(z)$ and $\Omega(z)$ now finally read
\begin{eqnarray}
~~~~~~~~~~~~~~~~~~~~~~~~~\psi(z) \approx \alpha \bigg[\dfrac{1}{3}+\dfrac{2}{3}z-\bigg(\dfrac{2}{9}+\dfrac{((1+a^{2})\beta +a \gamma)^{2}}{36r_{0}^{2}(1+a^{2}+a^{4})^{2}}\bigg)(1-z)^{2}\bigg]
\label{Psi_Taylor}
\end{eqnarray}
\begin{eqnarray}
\Phi(z) \approx \beta (1-z) + \dfrac{1}{3}\alpha^{2}\beta f(a)(1-z)^{2} 
\label{Phi_Taylor}
\end{eqnarray}
\begin{eqnarray}
\Omega(z) \approx \gamma (1-z) + \dfrac{1}{3}\alpha^{2}\gamma f(a)(1-z)^{2}
\label{Omega_Taylor}
\end{eqnarray}
where we have set $\alpha \equiv \psi(1)$, $\beta \equiv -\Phi^{\prime}(1)$ and $\gamma \equiv -\Omega^{\prime}(1)$~. 
\subsection{Solution near the asymptotic AdS region ($z = 0$)}
\noindent Near the AdS boundary $z \to 0$, the field $\psi(z)$ takes the following form~\cite{sg}
\begin{eqnarray}
\psi(z) \approx \dfrac{\langle \mathcal{O}_{\Delta} \rangle}{\sqrt{2}r_{0}^{\Delta}}z^{\Delta}
\label{Psi_z_0}
\end{eqnarray}
where $\Delta$ is the conformal dimension of the condensation operator $\mathcal{O}$ in the boundary field theory. In the present case, $\Delta = 1, 2$ which is a consequence of taking $m^{2} = -2$.\\
The fields $\Phi(z)$ and $\Omega(z)$ also admit similar  expansions around the asymptotic region 
\begin{eqnarray}
\Phi(z) \approx \mu - \dfrac{\rho}{r_{0}}z
\label{Phi_z_0}
\end{eqnarray}
\begin{eqnarray}
~~\Omega(z) \approx \nu - \dfrac{\zeta}{r_{0}}z~. 
\label{Omega_z_0}
\end{eqnarray}
\subsection{Matching and phase transition}
\noindent We now connect the approximate solution of the fields near the horizon to that of in the asymptotic regime. To do so we compare the first derivatives of eq.(s)(\ref{Phi_Taylor}, \ref{Phi_z_0}) at some point $z = z_{m}$
\begin{eqnarray}
\dfrac{\rho}{r_{0}}= \beta + \dfrac{2}{3}\alpha^{2}\beta f(a)(1-z_{m})~.
\label{rho_by_rnot}
\end{eqnarray}
Solving the above equation for $\alpha$ near $T \sim T_{c}$, we obtain
\begin{eqnarray}
\alpha = \sqrt{\dfrac{3}{f(a)(1-z_{m})}}\sqrt{1-\dfrac{T}{T_{c}}}
\label{alpha}
\end{eqnarray}
where we have used eq.(\ref{Hawking_temperature}) for $T$ and $T_{c}$ with $T_{c}$ being given by 
\begin{eqnarray}
T_{c}= \dfrac{3}{4\pi \varrho}\sqrt{\dfrac{\rho}{\tilde{\beta}}} ~ , ~ \tilde{\beta} = \dfrac{\beta}{r_{0}}~.
\label{Matching_T_c}
\end{eqnarray}
\subsection{Condensation operator with $\Delta = 1$}
\noindent For $\Delta = 1$, $\psi(z)$ can be written as 
\begin{eqnarray}
~~~~\psi(z) \approx \dfrac{\langle \mathcal{O}_{1} \rangle}{\sqrt{2}r_{0}}z
\label{Psi_z_for_delta_1}
\end{eqnarray}
\begin{eqnarray}
\Rightarrow \psi^{\prime}(z) \approx \dfrac{\langle \mathcal{O}_{1} \rangle}{\sqrt{2}r_{0}}~.
\label{Psi_prime_z_for_delta_1}
\end{eqnarray}
Matching eq.(\ref{Psi_Taylor}) with eq.(\ref{Psi_z_for_delta_1}) and derivative of eq.(\ref{Psi_Taylor}) with eq.(\ref{Psi_prime_z_for_delta_1}) at the intermediate point $z = z_{m}$,  we get 
\begin{eqnarray}
~~~~~\dfrac{\langle \mathcal{O}_{1} \rangle}{\sqrt{2}r_{0}}z_{m} =\alpha \bigg[\dfrac{1}{3}+\dfrac{2}{3}z_{m}-\bigg(\dfrac{2}{9}+\dfrac{((1+a^{2})\beta +a \gamma)^{2}}{36r_{0}^{2}(1+a^{2}+a^{4})^{2}}\bigg)(1-z_{m})^{2}\bigg] 
\label{Psi_matched}
\end{eqnarray}
\begin{eqnarray}
\dfrac{\langle \mathcal{O}_{1} \rangle}{\sqrt{2}r_{0}} =\alpha \bigg[\dfrac{2}{3}+\bigg(\dfrac{4}{9}+\dfrac{((1+a^{2})\beta +a \gamma)^{2}}{18r_{0}^{2}(1+a^{2}+a^{4})^{2}}\bigg)(1-z_{m})\bigg]~. 
\label{Psi_prime_matched}
\end{eqnarray}
Taking the ratio of eq.(s)(\ref{Psi_matched}) and (\ref{Psi_prime_matched}) and using eq.(\ref{Phi_Omega_prime_z_1}) to substitute for $\gamma \equiv -\Omega^{\prime}(1) $, we get 
\begin{eqnarray}
\tilde{\beta}=2\sqrt{\dfrac{(1+2z_{m}^{2})}{(1-z_{m}^{2})}}(1+a^{2})
\label{beta_tilde}
\end{eqnarray}
Substituting $\tilde{\beta}$ from eq.(\ref{beta_tilde}) in eq.(\ref{Matching_T_c}) yields the critical temperature to be of the form
\begin{eqnarray}
 T_{c}= \dfrac{3}{4\sqrt{2}\pi}\sqrt[4]{\bigg(\dfrac{1-z_{m}^{2}}{1+2z_{m}^{2}}\bigg)}\eta\sqrt{\rho}
\label{delta_1_T_c}
\end{eqnarray}
where  $\eta = \dfrac{1}{\varrho \sqrt{1+a^{2}}} = \dfrac{\Xi^{2}-a^{2}}{\Xi^{3/2}}$~. Taking derivative of $\eta (a)$ with respect to $a$, we find $$\dfrac{d\eta}{da} = \dfrac{(1+a^{2})^{\dfrac{3}{2}}(2a+4a^{3})-\dfrac{3}{2}(1+a^{2}+a^{4})(1+a^{2})^{\dfrac{1}{2}}}{(1+a^{2})^{3}}$$ which shows a minimum at $a \approx 0.5165$. Hence the critical temperature attains a minimum at $a \approx 0.5165$.\\
Also using eqs.(\ref{Phi_Omega_prime_z_1}, \ref{alpha}, \ref{beta_tilde}) in eq.(\ref{Psi_prime_matched}) near $T \sim T_{c}$, we find the value of the condensation to be 
\begin{eqnarray}
\dfrac{\langle \mathcal{O}_{1} \rangle}{T_{c}}=\dfrac{8\pi \varrho}{9}\sqrt{\dfrac{6}{f(a)(1-z_{m})}}\bigg(\dfrac{2+z_{m}}{1+z_{m}}\bigg)\sqrt{1-\dfrac{T}{T_{c}}}~.
\label{matching_Condensation_1}
\end{eqnarray}
From the above expression, we observe that the value of the critical temperature and the condensation depends on the rotation parameter of the black hole geometry. We find that there is a minimum in the critical temperature for the value of the rotation parameter $a \approx 0.5165$. This in turn implies that there is a range of $a$ from $[0,a_{min}]$ where the critical temperature decreases and hence superconductivity is not favoured. However, for the rotation parameter $a > a_{min}$, the crtical temperature increases indicating a situation favouring superconductivity. It is interesting to note that such an observation was made earlier in the literature in the context of Kerr black holes \cite{wald,KM}. There it was observed that higher rotation value of the black hole favours superconductivity by expelling magnetic field. We also observe that the value of the condensation operator decreases with the increase in the rotation parameter of the black hole. These observations have been presented in FIG.(s)(\autoref{figure(a)}), (\autoref{figure(b)}) where the matching method results have also been compared with the SL results.
\subsection{Condensation operator with $\Delta = 2$}
\noindent For $\Delta = 2$, we may write $\psi(z)$  as
\begin{eqnarray}
\psi(z) \approx \dfrac{\langle \mathcal{O}_{2} \rangle}{\sqrt{2}r_{0}^{2}}z^{2}
\label{Psi_z_for_delta_2}
\end{eqnarray}
\begin{eqnarray}
\psi^{\prime}(z) \approx \dfrac{\sqrt{2}\langle \mathcal{O}_{2} \rangle}{r_{0}^{2}}z ~ .
\label{Psi_prime_z_for_delta_2}
\end{eqnarray}
Now matching eq.(\ref{Psi_Taylor}) with eq.(\ref{Psi_z_for_delta_2}) and derivative of eq.(\ref{Psi_Taylor}) with eq.(\ref{Psi_prime_z_for_delta_2}) at the intermediate point $z = z_{m}$, we arrive at 
\begin{eqnarray}
~~~~~~~~~~~\dfrac{\langle \mathcal{O}_{2} \rangle}{\sqrt{2}r_{0}^{2}}z_{m}^{2} =\alpha \bigg[\dfrac{1}{3}+\dfrac{2}{3}z_{m}-\bigg(\dfrac{2}{9}+\dfrac{((1+a^{2})\beta +a \gamma)^{2}}{36r_{0}^{2}(1+a^{2}+a^{4})^{2}}\bigg)(1-z_{m})^{2}\bigg] 
\label{Psi_matched_2}
\end{eqnarray}
\begin{eqnarray}
\dfrac{\sqrt{2}\langle \mathcal{O}_{2} \rangle}{r_{0}^{2}}z_{m} =\alpha \bigg[\dfrac{2}{3}+\bigg(\dfrac{4}{9}+\dfrac{((1+a^{2})\beta +a \gamma)^{2}}{18r_{0}^{2}(1+a^{2}+a^{4})^{2}}\bigg)(1-z_{m})\bigg] ~ .
\label{Psi_prime_matched_2}
\end{eqnarray}
Taking ratio of eqs.(\ref{Psi_matched_2}) and (\ref{Psi_prime_matched_2}) and using eq.(\ref{Phi_Omega_prime_z_1}) to substitute for $\gamma$, we get 
\begin{eqnarray}
\tilde{\beta}=2\sqrt{\dfrac{(1+5z_{m})}{(1-z
_{m})}}(1+a^{2})~.
\label{tilde_beta_2}
\end{eqnarray}
Substituting $\tilde{\beta}$ from eq.(\ref{tilde_beta_2}) in eq.(\ref{Matching_T_c}), we get the following expression for the critical temperature
\begin{eqnarray}
 T_{c}= \dfrac{3}{4\sqrt{2}\pi}\sqrt[4]{\bigg(\dfrac{1-z_{m}}{1+5z_{m}}\bigg)}\eta\sqrt{\rho}~.
\label{delta_2_t_c}
\end{eqnarray}
%where  $\eta = \dfrac{1}{\varrho \sqrt{1+a^{2}}}$.\\
Once again using eq.(s)(\ref{Phi_Omega_prime_z_1}, \ref{alpha}, \ref{tilde_beta_2}) in eq.(\ref{Psi_prime_matched_2}) near $T \sim T_{c}$, the condensation value is found to be 
\begin{eqnarray}
\dfrac{\sqrt{\langle \mathcal{O}_{2} \rangle}}{T_{c}}=\dfrac{4\pi \varrho}{3}\sqrt[4]{\dfrac{2(2+z_{m})^{2}}{3f(a)(1-z_{m})}}\sqrt[4]{1-\dfrac{T}{T_{c}}}
\label{matching_condensation_2}
\end{eqnarray}
which once again exhibits the effect of the rotation parameter of the black hole geometry. The dependence of $T_{c}$ and the condensation operator on the rotation parameter has been displayed in FIG.(\autoref{figure(a)}). \\
In the table below, we have presented the results for the critical temperature $T_{c}$ for different choices of the matching points $z_{m}$ for the two possible values of the condensation operators. It can be seen that the results agree with the SL results given in \cite{la}.

\begin{table}[h]
\begin{center}
\begin{tabular}{ |c| c| c|}
\hline
~~~~~~Matching point, $z_{m}$~~~~~~ & \multicolumn{2}{c|}{~~~$T_{c}$ from matching method~~~~~~}         \\
\hline
& ~~~~~~$\Delta = 1$~~~~~~ & $\Delta = 2$ ~~~\\
\hline 
0.1 & 0.1675$\eta\sqrt{\rho}$ & 0.1486$\eta\sqrt{\rho}$\\
\hline
0.3 & 0.1582$\eta\sqrt{\rho}$ & 0.1228$\eta\sqrt{\rho}$\\
\hline 
0.5 & 0.1419$\eta\sqrt{\rho}$ & 0.1038$\eta\sqrt{\rho}$\\
\hline
0.7 & 0.1203$\eta\sqrt{\rho}$ & 0.0858$\eta\sqrt{\rho}$\\
\hline 
\end{tabular}
\label{tab1}
\end{center}
\caption{Critical temperature at different matching points for $\Delta = 1, 2$.}
\end{table}
\section{St{\"u}rm-Liouville Analysis}
\noindent \hypertarget{sec4}{In} this section, we carry out the analysis using a variational approach, also known as the St{\"u}rm-Liouville eigenvalue method and compare our results with those obtained from the matching method discussed earlier.\\
To proceed further, we first recall that the scalar field $\psi(z)$, vanishes at the critical temperature $T_{c}$. Hence, at $T = T_{c}$ eq.(s)(\ref{Phi_field_eq_2}, \ref{Omega_field_eq_2}) simplify to the following equations  
\begin{eqnarray}
\Phi^{\prime\prime}+\dfrac{3a^{2}z^{2}}{(1+a^{2}+a^{4})(1-z^{3})}\Phi^{\prime} + \dfrac{3a(1+a^2)z^{2}}{(1+a^{2}+a^{4})(1-z^{3})}\Omega^{\prime}=0
\label{Phi_field_at_T_c}
\end{eqnarray}
\begin{eqnarray}
\Omega^{\prime\prime}-\dfrac{3(1+a^{2})^{2}z^{2}}{(1+a^{2}+a^{4})(1-z^{3})}\Omega^{\prime} - \dfrac{3a(1+a^{2})z^{2}}{(1+a^{2}+a^{4})(1-z^{3})}\Phi^{\prime}=0 ~ .
\label{Omega_field_at_T_c}
\end{eqnarray}
Notice that eq.(s)(\ref{Phi_field_at_T_c}, \ref{Omega_field_at_T_c}) are still coupled differential equations for $\Phi(z)$ and $\Omega(z)$. However, we can decouple the above eq.(s) in the following form 
\begin{eqnarray}
\Phi^{\prime\prime\prime} + \dfrac{2}{z}\bigg(\dfrac{2z^{3}+1}{z^{3}-1}\bigg)\Phi^{\prime\prime}=0
\label{decoupled_phi}
\end{eqnarray}
\begin{eqnarray}
\Omega^{\prime\prime\prime} + \dfrac{2}{z}\bigg(\dfrac{2z^{3}+1}{z^{3}-1}\bigg)\Omega^{\prime\prime}=0 ~ .
\label{decoupled_omega}
\end{eqnarray}
Now the decoupled eq.(s)(\ref{decoupled_phi}, \ref{decoupled_omega}) yield the following exact solutions for $\Phi(z)$ and $\Omega(z)$ \cite{la}.
\begin{eqnarray}
\Phi(z)= \mu -\dfrac{\rho}{r_{0c}}z+C_{1}\bigg[\sqrt{12}\arctan\bigg(\dfrac{1+2z}{\sqrt{3}}\bigg)+\ln \bigg(\dfrac{1+z+z^{2}}{(1-z)^{2}}\bigg)\bigg]
\label{exact_Phi}
\end{eqnarray}
\begin{eqnarray}
\Omega(z)= \nu -\dfrac{\zeta}{r_{0c}}z+C_{2}\bigg[\sqrt{12}\arctan\bigg(\dfrac{1+2z}{\sqrt{3}}\bigg)+\ln \bigg(\dfrac{1+z+z^{2}}{(1-z)^{2}}\bigg)\bigg]
\label{exact_Omega}
\end{eqnarray}
where $\mu$, $\rho$, $\nu$ and $\zeta$ have the usual interpretations in the boundary field theory from the AdS/CFT dictionary. Now in view of the boundary conditions at the horizon $\Phi(1)=0$ and $\Omega(1)=0$, we  write 
\begin{eqnarray}
\Phi=\lambda r_{0c}(1-z)
\label{Phi_lambda_relation}
\end{eqnarray}
\begin{eqnarray}
\Omega=\bar{\lambda} r_{0c}(1-z)
\label{Omega_lambda_relation}
\end{eqnarray}
where $\lambda=\dfrac{\rho}{r_{0c}^{2}},~ r_{0c}$ is the horizon radius at the critical temperature and $\bar{\lambda}$ is related to $\lambda$ through following relation
\begin{eqnarray}
a\lambda + (1+a^{2})\bar{\lambda}=0 
\label{lamda_lamda_bar_relation}
\end{eqnarray}
which readily follows from using eq.(s)(\ref{Phi_Omega_prime_z_1}, \ref{Phi_lambda_relation}, \ref{Omega_lambda_relation}).\\
We now consider the following non-trivial form of the field $\psi(z)$ \cite{st}
\begin{eqnarray}
\psi(z)=\dfrac{\langle \mathcal{O}_{\Delta} \rangle}{\sqrt{2}r_{0}^{\Delta}}z^{\Delta}F(z)
\label{psi_sturm}
\end{eqnarray}
such that $F(0)=1$ and $F^{\prime}(0)=0$. With this form of the field $\psi(z)$, eq.(\ref{Psi_field_eq_2}) becomes
\begin{eqnarray}
-F^{\prime\prime}+\dfrac{1}{z}\bigg(\dfrac{2+z^{3}}{1-z^{3}}-2\Delta \bigg)F^{\prime} + \dfrac{\Delta^{2}z}{(1-z^{3})}F=\dfrac{{\tilde{\lambda}}^{2}}{(1+z+z^{2})^{2}}F
\label{field_eq_in_F_z}
\end{eqnarray}
where $\tilde{\lambda}=\dfrac{\lambda}{(1+a^{2})}$~. 
\subsection{Analysis for $\Delta=1$}
\noindent For $\Delta=1$, eq.(\ref{field_eq_in_F_z}) takes the following form
\begin{eqnarray}
-F^{\prime\prime}+\dfrac{3z^{2}}{(1-z^{3})}F^{\prime}+\dfrac{z}{(1-z^{3})}F=\dfrac{\tilde{\lambda}^{2}}{(1+z+z^{2})^{2}}F
\label{field_eq_F_z_delta_1}
\end{eqnarray}
which can be rearranged into the St{\"u}rm-Liouville form 
\begin{eqnarray}
\dfrac{d(p(z)F^{\prime})}{dz}+q(z)F+\tilde{\lambda}^{2}r(z)F=0
\label{Sturm_form_F_z_eq_delta_1}
\end{eqnarray}
with $$p(z) = (1-z^{3}), ~ q(z) = -z, ~ r(z) = \bigg(\dfrac{1-z}{1+z+z^{2}}\bigg)~. $$ 
Now the eigenvalue $\tilde{\lambda}^{2}$ which minimizes the above expression is given as
\begin{eqnarray}
\tilde{\lambda}^{2} = \dfrac{\displaystyle\int\limits_{0}^{1}dz \big(p(z)F^{\prime 2} - q(z)F^{2}\big)}{\displaystyle\int\limits_{0}^{1}dzr(z)F^{2}} = \dfrac{\displaystyle\int\limits_{0}^{1}dz \big((1-z^{3})F^{\prime 2} + zF^{2}\big)}{\displaystyle\int\limits_{0}^{1}dz\dfrac{(1-z)}{(1+z+z^{2})}F^{2}}~. 
\label{eigenvalue_integral_delta_1}
\end{eqnarray}
To proceed further, we take the trial function as \cite{st} 
 \begin{eqnarray}
 F_{\tilde{\alpha}}(z)=(1-\tilde{\alpha} z^{2}) 
 \label{trail_function}
 \end{eqnarray}
which satisfies the boundary conditions $F_{\tilde{\alpha}}(0)=1$ and $F_{\tilde{\alpha}}^{\prime}(0)=0$. One should notice that this choice of the trial function is also compatible with the boundary conditions for the scalar field $\psi(z)$. Now with this trial function eq.(\ref{eigenvalue_integral_delta_1}) takes the following form
$$\tilde{\lambda}_{\tilde{\alpha}}^{2}=\dfrac{\displaystyle\int\limits_{0}^{1}dz \big(4\tilde{\alpha}^{2}z^{2}(1-z^{3}) + z(1-\tilde{\alpha} z^{2})^{2}\big)}{\displaystyle\int\limits_{0}^{1}dz\dfrac{(1-z)}{(1+z+z^{2})}(1-\tilde{\alpha} z^{2})^{2}}$$
\begin{eqnarray}
~~~~~~~~~~~~~~~~~~~~~~~~~~~~~~~= \dfrac{6-6\tilde{\alpha}+10\tilde{\alpha}^{2}}{2\sqrt{3}\pi -6 ln3 + 4(\sqrt{3}\pi + 3 ln3 - 9)\tilde{\alpha} + (12 ln3 - 13)\tilde{\alpha}^{2}}~.
\label{eigenvalue_integral_f_alpha_delta_1}
\end{eqnarray}
From the above expression, one can see that $\tilde{\lambda}_{\tilde{\alpha}}^{2}$ attains minima with respect to $\tilde{\alpha}$ at $\tilde{\alpha} \approx 0.2389$. This yields
$$\tilde{\lambda}^{2}_{0.2389} \approx 1.268 ~ .$$
Using eq.(\ref{Hawking_temperature}), the critical temperature is given by the following relation 
$$ T_{c} = \dfrac{3r_{0c}}{4\pi\varrho} =  \dfrac{3}{4\pi\varrho} \sqrt{\dfrac{\rho}{\lambda}}   $$

\begin{eqnarray}
~~~~~~~~~~~~~~~~~~~~~~~= \dfrac{3}{4\pi\varrho} \sqrt{\dfrac{\rho}{\tilde{\lambda}}} \dfrac{1}{\sqrt{1+a^{2}}} =\dfrac{3}{4\pi}\eta \sqrt{\dfrac{\rho}{\tilde{\lambda}}}
\label{T_c_eq_Lin_Abdalla}
\end{eqnarray}
where $\eta = \dfrac{1}{\varrho\sqrt{1+a^{2}}}$.\\
Now replacing $\tilde{\lambda}$ in the above relation with the estimated value of $\tilde{\lambda}_{0.2389}$ obtained through the SL analysis shown above, we finally get the critical temperature to be
\begin{eqnarray}
T_{c}=0.225 \eta \sqrt{\rho}~.
\label{Sturm-Liouville_critical_temp_delta_1}
\end{eqnarray}
Once again we notice that the rotation parameter affects the critical temperature. It is reassuring to note that in the no rotation limit that is $a = 0 $, we get the critical temperature for the static black hole case \cite{sg}.\\
We now move ahead to find the value of condensation operator utilizing this analysis. For this purpose we first write the following approximate expressions for the fields $\psi(z)$, $\Phi(z)$ and $\Omega(z)$ near the critical temperature $T_{c}$
\begin{eqnarray}
\psi(z)=\dfrac{\langle \mathcal{O}_{1} \rangle}{\sqrt{2}r_{0}}zF(z)
\label{psi_with_fluctuation_delta_1}
\end{eqnarray}
\begin{eqnarray}
~~~~~~~~~~~~\dfrac{\Phi(z)}{r_{0}} \approx \lambda (1-z)+\dfrac{\langle \mathcal{O}_{1} \rangle ^{2}}{r_{0}^{2}} \chi(z)
\label{Phi_with_fluctuation_delta_1}
\end{eqnarray}
\begin{eqnarray}
~~~~~~~~~~~~\dfrac{\Omega(z)}{r_{0}} \approx \bar{\lambda} (1-z)+\dfrac{\langle \mathcal{O}_{1} \rangle ^{2}}{r_{0}^{2}} \varsigma(z)
\label{Omega_with_fluctuation_delta_1}
\end{eqnarray}
where $\chi(z)$ and $\varsigma(z)$ are fluctuation fields.\\
Now using eq.(s)(\ref{psi_with_fluctuation_delta_1}, \ref{Phi_with_fluctuation_delta_1},  \ref{Omega_with_fluctuation_delta_1}) in eq.(s)(\ref{Phi_field_eq_2},  \ref{Omega_field_eq_2}),  we get the following field eq.(s) for the fluctuation fields $\chi(z)$  and $\varsigma(z)$ to be
\begin{eqnarray}
\chi^{\prime\prime}+\dfrac{3az^{2}\Lambda^{\prime}}{(1+a^{2}+a^{4})(1-z^{3})}= \dfrac{\lambda(1-z)F^{2}}{(1-z^{3})}
\label{chi_eq_delta_1}
\end{eqnarray}
\begin{eqnarray}
\varsigma^{\prime\prime}+\dfrac{3(1+a^{2})z^{2}\Lambda^{\prime}}{(1+a^{2}+a^{4})(z^{3}-1)}= \dfrac{\bar{\lambda}(1-z)F^{2}}{(1-z^{3})}
\label{var_sigma_eq_delta_1}
\end{eqnarray}
where $\Lambda(z)$ is given by  $$\Lambda(z)=a\chi(z) + (1+a^{2})\varsigma(z)~.$$
Now multiplying eq.(\ref{chi_eq_delta_1}) by $a$ and eq.(\ref{var_sigma_eq_delta_1}) by ($1 + a^2$) and adding them up together with the condition (\ref{lamda_lamda_bar_relation}), we arrive at the following equation that is entirely given in terms of $\Lambda(z)$
\begin{eqnarray}
\Lambda^{\prime\prime}+\dfrac{3z^{2}}{(z^{3}-1)}\Lambda^{\prime}=0~.
\label{Lambda_eq}
\end{eqnarray}
With the boundary conditions discussed earlier, we can solve eq.({\ref{Lambda_eq}}) to obtain
\begin{eqnarray}
\Lambda^{\prime}(z)=0~.
\label{Lambda_prime_z}
\end{eqnarray}
With this condition eq.(\ref{chi_eq_delta_1}) reduces to
\begin{eqnarray}
\chi^{\prime\prime}=\dfrac{\lambda(1-z)}{(1-z^{3})}F^{2}~.
\label{Chi_prime_prime_delta_1}
\end{eqnarray}
Integrating eq.(\ref{Chi_prime_prime_delta_1}) with respect to $z$ between 0 and 1, we get the following result
\begin{eqnarray}
\chi^{\prime}(1)-\chi^{\prime}(0)= \lambda \displaystyle\int\limits_{0}^{1}dz \dfrac{1-z}{1-z^{3}}F^{2}~.
\label{chi_prime_0_1_delta_1}
\end{eqnarray}
We now expand the fluctuation field $\chi(z)$ about $z = 0$ \begin{eqnarray}
\chi(z)=\chi(0)+\chi^{\prime}(0)z+... ~.
\label{Chi_z_around_0}
\end{eqnarray}  
Using the above expansion of $\chi(z)$ in eq.(\ref{Phi_with_fluctuation_delta_1}) and equating it to the asymptotic solution for $\Phi(z)$ given in eq.(\ref{Phi_z_0}), we get
\begin{eqnarray}
\dfrac{\mu}{r_{0}}-\dfrac{\rho}{r_{0}^{2}}z=\lambda(1-z)+\dfrac{\langle \mathcal{O}_{1} \rangle ^{2}}{r_{0}^{2}}( \chi(0)+\chi^{\prime}(0)z+... )~. 
\label{Operator_1_prelims}
\end{eqnarray}
Comparing the coefficients of $z$ on both sides of the above equation gives
\begin{eqnarray}
\dfrac{\rho}{r_{0}^{2}}=\lambda-\dfrac{\langle \mathcal{O}_{1} \rangle ^{2}}{r_{0}^{2}}\chi^{\prime}(0)~.
\label{Operator_1_and_rho_relation}
\end{eqnarray} 
Now replacing $\chi^{\prime}(z)$ from eq.(\ref{chi_prime_0_1_delta_1}) in eq.(\ref{Operator_1_and_rho_relation}), we obtain
\begin{eqnarray}
\dfrac{\rho}{r_{0}^{2}}=\lambda\bigg(1+\dfrac{\langle \mathcal{O}_{1} \rangle ^{2}}{r_{0}^{2}}   \mathcal{A}\bigg)-\dfrac{\langle \mathcal{O}_{1} \rangle ^{2}}{r_{0}^{2}}\chi^{\prime}(1)
\label{Operator_1_and_chi_prime_1_relation}
\end{eqnarray}
where ~ $\mathcal{A}=\displaystyle\int\limits_{0}^{1}dz \dfrac{1-z}{1-z^{3}}F^{2}$~.
To proceed further we need the value of $\chi^{\prime}(1)$ in the above equation. \\
To obtain this we need to recall that we have earlier found the exact solutions for the fields $\Phi(z)$ and $\Omega(z)$ at $T = T_{c}$ in eq.(s)(\ref{Phi_lambda_relation}, \ref{Omega_lambda_relation}). 
We can assume that near $T_{c}$, the field $\Phi(z)$ takes the following form 
\begin{eqnarray}
\Phi=\lambda r_{0}(1-z)+ \dfrac{{\langle \mathcal{O}_{1} \rangle}^{2}}{r_{0}}\bigg(\chi(1)+(z-1)\chi^{\prime}(1)+...\bigg)~.
\label{Phi_chi_1_relation_new}
\end{eqnarray}
We now equate the coefficient of ($1-z$) in above equation with that of the solution for the field $\Phi(z)$ at $T_{c}$ given by eq.(\ref{Phi_lambda_relation})
\begin{eqnarray}
\lambda r_{0c}= \lambda r_{0} - \dfrac{{\langle \mathcal{O}_{1} \rangle}^{2}}{r_{0}}\chi^{\prime}(1)~.
\label{lambda_chi_prime_1_operator_1}
\end{eqnarray}
Rearranging the above equation, we have 
\begin{eqnarray}
\chi^{\prime}(1)=\dfrac{\lambda r_{0}}{{\langle \mathcal{O}_{1} \rangle}^{2}}(r_{0}-r_{0c})~.
\label{chi_prime_1_delta_1}
\end{eqnarray}
Using the above relation in eq.(\ref{Operator_1_and_chi_prime_1_relation}), we obtain
\begin{eqnarray}
\dfrac{\rho}{r_{0}^{2}}=\lambda\bigg(\dfrac{r_{0c}}{r_{0}}+\dfrac{\langle \mathcal{O}_{1} \rangle ^{2}}{r_{0}^{2}}   \mathcal{A}\bigg)~.
\label{result_rho_lambda_delta_1}
\end{eqnarray} 
Replacing $r_{0}, \rho$ and $\lambda$ using eq.(s)(\ref{Hawking_temperature}, \ref{T_c_eq_Lin_Abdalla}) finally leads to the following result 
\begin{eqnarray}
\langle \mathcal{O}_{1} \rangle = \gamma T_{c}\sqrt{\bigg(1-\dfrac{T}{T_{c}}\bigg)}
\label{O_1_T_c_relation_2}
\end{eqnarray}
$$\gamma = \dfrac{4\pi}{3}\bigg(\dfrac{1+a^{2}}{1+a^{2}+a^{4}}\bigg)\sqrt{\dfrac{1}{\mathcal{A}}}~.$$
Once again we notice that the rotation parameter $a$ affects the condensate.
\subsection{Analysis for $\Delta=2$}
\noindent We now carry out the analysis for $\Delta=2$.
For $\Delta=2$, eq.(\ref{field_eq_in_F_z}) reduces to
\begin{eqnarray}
-F^{\prime\prime}+\dfrac{1}{z}\bigg(\dfrac{5z^{3}-2}{1-z^{3}}\bigg)F^{\prime}+\dfrac{4z}{(1-z^{3})}F=\dfrac{\tilde{\lambda}^{2}}{(1+z+z^{2})^{2}}F
\label{delta_2_field_F_z}
\end{eqnarray}
which we can recast into the St{\"u}rm-Liouville form as 
\begin{eqnarray}
\dfrac{d(1-z^{3})F^{\prime}z^{2}}{dz}-4z^{3}F+\dfrac{\tilde{\lambda}^{2}z^{2}(1-z)}{(1+z+z^{2})}F=0~.
\label{sturm_liouville_form_F_z_delta_2}
\end{eqnarray}
The eigenvalue $\tilde{\lambda}^{2}$ in this case reads
\begin{eqnarray}
\tilde{\lambda}^{2}=\dfrac{\displaystyle\int\limits_{0}^{1}dz \big(z^{2}(1-z^{3})F^{\prime 2} +4z^{3}F^{2}\big)}{\displaystyle\int\limits_{0}^{1}dz\dfrac{z^{2}(1-z)}{(1+z+z^{2})}F^{2}}~.
\label{eigenvalue_integral_delta_2}
\end{eqnarray}
Once again we take the previous trial function (\ref{trail_function}) as an ansatz to estimate the eigenvalue.\\
With this ansatz we can write eq.(\ref{eigenvalue_integral_delta_2}) as
$$\tilde{\lambda}_{\tilde{\alpha}}^{2}=\dfrac{\displaystyle\int\limits_{0}^{1}dz \big(4\tilde{\alpha}^{2}(z^{4}-z^{7}) + 4z^{3}(1-\tilde{\alpha} z^{2})^{2}\big)}{\displaystyle\int\limits_{0}^{1}dz\dfrac{z^{2}(1-z)}{(1+z+z^{2})}(1-\tilde{\alpha} z^{2})^{2}}$$
\begin{eqnarray}
~~~~~~~~= \dfrac{60-80\tilde{\alpha}+48\tilde{\alpha}^{2}}{2.62765 -(1.8335)\tilde{\alpha} + (0.45561)\tilde{\alpha}^{2}}~.
\label{solved_eigenvalue_delta_2}
\end{eqnarray}
Eq.(\ref{solved_eigenvalue_delta_2}) attains its minima for $\tilde{\alpha} \approx 0.60159$. This gives
$$\tilde{\lambda}^{2}_{0.60159} \approx 17.309~.$$
Hence the critical temperature, in this case, is given by the following relation  
$$ ~~~T_{c} = \dfrac{3r_{0c}}{4\pi\varrho} =  \dfrac{3}{4\pi\varrho} \sqrt{\dfrac{\rho}{\lambda}}$$ 
$$~~~~~~~~~~~~~~~~~~~~~~~= \dfrac{3}{4\pi\varrho} \sqrt{\dfrac{\rho}{\tilde{\lambda}}} \dfrac{1}{\sqrt{1+a^{2}}} =\dfrac{3}{4\pi}\eta \sqrt{\dfrac{\rho}{\tilde{\lambda}}}$$
%\begin{eqnarray}
%T_{c}=\dfrac{3}{4\pi}\eta \sqrt{\dfrac{\rho}{\tilde{\lambda}}}
%\label{84}
%\end{eqnarray}
\begin{eqnarray}
~~=0.117 \eta \sqrt{\rho}
\label{T_c_delta_2}
\end{eqnarray}
where we have used the value of $\tilde{\lambda}^{2}$ given above. \\
As in the previous case, we now consider the following forms for the fields $\psi(z), \Phi(z)$ and $\Omega(z)$ 
\begin{eqnarray}
\psi(z) = \dfrac{\langle \mathcal{O}_{2} \rangle}{\sqrt{2}r_{0}^{2}}z^{2}F(z)
\label{psi_z_delta_2}
\end{eqnarray}
\begin{eqnarray}
~~~~~~~~~~~~\dfrac{\Phi(z)}{r_{0}} \approx \lambda (1-z)+\dfrac{\langle \mathcal{O}_{2} \rangle ^{2}}{r_{0}^{4}} \chi(z)
\label{Phi_z_delta_2}
\end{eqnarray}
\begin{eqnarray}
~~~~~~~~~~~~\dfrac{\Omega(z)}{r_{0}} \approx \bar{\lambda} (1-z)+\dfrac{\langle \mathcal{O}_{2} \rangle ^{4}}{r_{0}^{4}} \varsigma(z)~.
\label{Omega_z_delta_2}
\end{eqnarray}
Using eq.(s)( \ref{psi_z_delta_2}, \ref{Phi_z_delta_2}, \ref{Omega_z_delta_2}) in eq.(s)(\ref{Phi_field_eq_2}, \ref{Omega_field_eq_2}) we get the following equations in terms of the fluctuation fields
\begin{eqnarray}
\chi^{\prime\prime}+\dfrac{3az^{2}\Lambda^{\prime}}{(1+a^{2}+a^{4})(1-z^{3})}= \dfrac{\lambda(1-z)z^{2}F^{2}}{(1-z^{3})}
\label{chi_field_eq_delta_2}
\end{eqnarray}
\begin{eqnarray}
\varsigma^{\prime\prime}+\dfrac{3(1+a^{2})z^{2}\Lambda^{\prime}}{(1+a^{2}+a^{4})(z^{3}-1)}= \dfrac{\bar{\lambda}(1-z)z^{2}F^{2}}{(1-z^{3})}
\label{varsigma_field_eq_delta_2}
\end{eqnarray}
where ~ $\Lambda(z)=a\chi(z) + (1+a^{2})\varsigma(z)$~.\\
Now multiplying eq.(\ref{chi_field_eq_delta_2}) by $a$ and eq.(\ref{varsigma_field_eq_delta_2}) by ($1 + a^2$) and adding them up together with the boundary condition given in eq.(\ref{lamda_lamda_bar_relation}), we arrive at the following equation
\begin{eqnarray}
\Lambda^{\prime\prime}+\dfrac{3z^{2}}{(z^{3}-1)}\Lambda^{\prime}=0~.
\label{delta_2_lambda_eq}
\end{eqnarray}
Solving the above equation yields
\begin{eqnarray}
\Lambda^{\prime}(z)=0.
\label{lamda_prime_delta_2}
\end{eqnarray}
Using above condition in eq.(\ref{chi_field_eq_delta_2}) gives the following equation 
\begin{eqnarray}
\chi^{\prime\prime}=\dfrac{\lambda(1-z)}{(1-z^{3})}z^{2}F(z)^{2}~.
\label{chi_prime_prime_eq_delta_2}
\end{eqnarray}
Integrating this equation with respect to $z$ between 0 and 1, we arrive at following result 
\begin{eqnarray}
\chi^{\prime}(1)-\chi^{\prime}(0)= \lambda \displaystyle\int\limits_{0}^{1}dz \dfrac{1-z}{1-z^{3}}z^{2}F(z)^{2}~.
\label{chi_prime_0_1_delta_2}
\end{eqnarray}
 Now we expand the fluctuation field $\chi(z)$ about $z = 0$ as
\begin{eqnarray}
\chi(z)=\chi(0)+\chi^{\prime}(0)z+...
\label{chi_about_0_delta_2}
\end{eqnarray}  
Now we use above expansion in eq.(\ref{Phi_z_delta_2}) and equate it to the asymptotic solution for $\Phi(z)$ given in  eq.(\ref{Phi_z_0}) to get
\begin{eqnarray}
\dfrac{\mu}{r_{0}}-\dfrac{\rho}{r_{0}^{2}}z=\lambda(1-z)+\dfrac{\langle \mathcal{O}_{2} \rangle ^{2}}{r_{0}^{4}}( \chi(0)+\chi^{\prime}(0)z+... )~. 
\label{rho_O_2_relation_1}
\end{eqnarray}
Comparing the coefficients of $z$ on both sides of the above equation, we get
\begin{eqnarray}
\dfrac{\rho}{r_{0}^{2}}=\lambda-\dfrac{\langle \mathcal{O}_{2} \rangle ^{2}}{r_{0}^{4}}\chi^{\prime}(0)~.
\label{rho_O_2_chi_0}
\end{eqnarray} 
Replacing $\chi^{\prime}(0)$ from eq.(\ref{chi_prime_0_1_delta_2}) in the above relation, we get
\begin{eqnarray}
\dfrac{\rho}{r_{0}^{2}}=\lambda\bigg(1+\dfrac{\langle \mathcal{O}_{2} \rangle ^{2}}{r_{0}^{4}}   \mathcal{B}\bigg)-\dfrac{\langle \mathcal{O}_{2} \rangle ^{2}}{r_{0}^{4}}\chi^{\prime}(1)
\label{rho_O_2_chi_1}
\end{eqnarray}
where~ $\mathcal{B}=\displaystyle\int\limits_{0}^{1}dz \dfrac{1-z}{1-z^{3}}z^{2}F(z)^{2}$~.
 
\noindent To estimate $\chi^{\prime}(1)$, once again we can use the exact solutions for the fields at $T = T_{c}$ given by eq.(s)(\ref{Phi_lambda_relation}, \ref{Omega_lambda_relation}) and $\psi(z)=0$.
Also, near $T_{c}$ we can write the following expression for the field $\Phi(z)$ 
\begin{eqnarray}
\Phi=\lambda r_{0}(1-z)+ \dfrac{{\langle \mathcal{O}_{2} \rangle}^{2}}{r_{0}^{3}}\bigg(\chi(1)+(z-1)\chi^{\prime}(1)+...\bigg)~.
\label{Phi_near_T_c_O_2}
\end{eqnarray}
Equating the coefficients of ($1-z$) in the above equation with that of the solution at $T = T_{c}$ given in eq.(\ref{Phi_lambda_relation}) yields
\begin{eqnarray}
\lambda r_{0c}= \lambda r_{0} - \dfrac{{\langle \mathcal{O}_{2} \rangle}^{2}}{r_{0}^{3}}\chi^{\prime}(1)~.
\label{lambda_chi_prime_1_O_2}
\end{eqnarray}
Rearranging the above expression, we find
\begin{eqnarray}
 \chi^{\prime}(1)=\dfrac{\lambda r_{0}^{3}}{{\langle \mathcal{O}_{2} \rangle}^{2}}(r_{0}-r_{0c})~.
\label{chi_prime_1_delta_2}
\end{eqnarray}
Now we use the above relation for $\chi^{\prime}(1)$ in eq.(\ref{lambda_chi_prime_1_O_2}) to arrive at the following result
\begin{eqnarray}
\dfrac{\rho}{r_{0}^{2}}=\lambda\bigg(\dfrac{r_{0c}}{r_{0}}+\dfrac{\langle \mathcal{O}_{2} \rangle ^{2}}{r_{0}^{4}}   \mathcal{B}\bigg)~.
\label{105}
\end{eqnarray} 
\hypertarget{sec5}{Finally} rewriting $\rho, r_{0}$ and $r_{0c}$ in terms of $T$ and $T_{c}$, we arrive at the following result 
\begin{eqnarray}
\sqrt{\langle \mathcal{O}_{2} \rangle} = \gamma T_{c}\sqrt[4]{\bigg(1-\dfrac{T}{T_{c}}\bigg)}
\label{result_O_2}
\end{eqnarray}
 $$\gamma = \dfrac{4\pi}{3}\bigg(\dfrac{1+a^{2}}{1+a^{2}+a^{4}}\bigg)\sqrt[4]{\dfrac{1}{\mathcal{B}}}~.$$
\section{Conclusions}
\begin{figure}[h!]
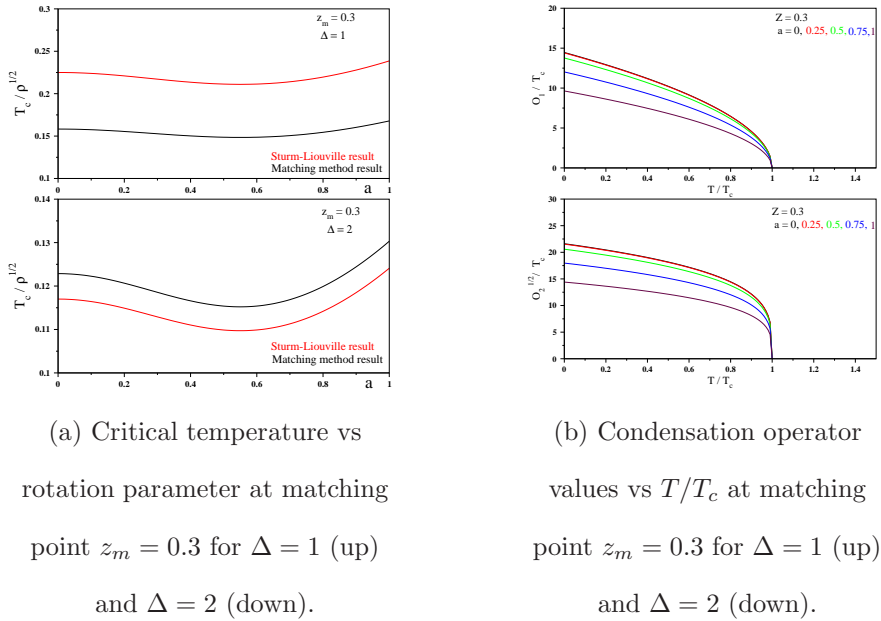

 \subfloat[Critical temperature vs rotation parameter at matching point $z_{m} = 0.3$ for $\Delta = 1$ (up) and $\Delta = 2$ (down).\label{figure(a)}]{{\includegraphics[scale=0.255]{match_sturm.eps} }}
\qquad
~~~~~~
\subfloat[Condensation operator values vs $T / T_{c}$ at matching point $z_{m}=0.3$ for $\Delta = 1$ (up) and $\Delta = 2$ (down).\label{figure(b)}]{{\includegraphics[scale=0.24]{O1_Tc.eps} }}
\qquad
\caption{Analytical results using matching method and St{\"u}rm-Liouville analysis}
\end{figure}

\noindent We now summarize our findings. In this paper we have analytically investigated a model for the rotating holographic $s$-wave superconductor in the probe limit. We have calculated the critical temperature and the condensation operator values for the two possible conformal dimensions $\Delta = 1, 2$ using matching method as well as St{\"u}rm-Liouville eigenvalue analysis. From our investigation we notice that if we increase the rotation parameter, $a$, of the black hole, the critical temperature first decreases and thereafter it again starts to rise from the value of $a \approx 0.5165$. This behaviour of the critical temperature obtained using the matching method as well as the St{\"u}rm-Liouville analysis, for $\Delta = 1, 2$, is shown in FIG.(\autoref{figure(a)}). From the Figure, it is evident that there is a value of the rotation at which the critical temperature is a minimum which indicates that the superconductivity is not favoured for this value of the rotation parameter of the black hole. This is similar to the observations made in \cite{js} where it was observed that below $T_{c}$, the transition temperature at zero rotation, there exists a critical value of the rotation which breaks the superconductivity order. In FIG.(\autoref{figure(b)}), we find that the condensation operator values show a second order phase transition but keep on decreasing with increase in the value of the rotation parameter of the black hole. For higher values of the rotation parameter, value of the condensation operators falls sharply. However, it is interesting to note that for rapidly rotating black holes ($a \rightarrow 1$), a small amount of condensate forms with a higher value of the critical temperature than their non-rotating counterpart as can be clearly seen in FIG.(\autoref{figure(a)}) and (\autoref{figure(b)}). This is interesting since such an observation was made long back where it was demonstrated that higher values of spin is favourable for superconductivity \cite{wald, KM}.

\noindent In FIG.(\autoref{figure(b)}, (up) and (down)), upper most curve represents values of the condensation operator for non-rotating black hole, that is $a = 0$, and subsequently lower curves are associated with increasing values of the rotation parameter of the black hole such that the lowest one corresponds to the maximally rotating black hole with the rotation parameter $a = 1$. \\

\noindent {\bf{Acknowledgements}}:  SG acknowledges the support by DST SERB under Start Up Research Grant (Young Scientist), File
No.YSS/2014/000180. SG also acknowledge the Visiting Associateship at IUCAA, Pune. The authors would also like to thank the referee for useful comments.

%%%%%%%%%%%%%%%%%%%%%%%%%%%%%%%%%%%%%%%%%%%%%%%%%%%%%%%%%%%%%%%%%%%%%%%%%%%

%%%%%%%%%%%%%%%%%%%%%%%%%%%%%%%%%%%%%%%%%%%%%%%%%%%%%%%%%%%%%%%%%%%%%%%%%%%%
%%%%%%%%%%%%%%%%%%%%%%%%%%%%%%%%%%%%%%%%%%%%%%%%%%%%%%%%%%%%%%%%%%%%%%%%%%%%

\begin{thebibliography}{99}


\bibitem{bm}J.G. Bednorz, K.A. M{\"u}ller, \href{https://doi.org/10.1007/BF01303701}{Z. Physik B - Condensed Matter (1986) 64: 189.}

\bibitem{jm}J. Maldacena, \href{https://doi.org/10.1023/A:1026654312961}{International Journal of Theoretical Physics (1999) 38: 1113.}

\bibitem{SSG}S. S. Gubser, \href{https://doi.org/10.1103/PhysRevD.78.065034}{Phys. Rev. D 78, 065034}


\bibitem{hhh}S. A. Hartnoll, C. P. Herzog, G. T. Horowitz, \href{https://doi.org/10.1103/PhysRevLett.101.031601}{Phys. Rev. Lett. 101, 031601.}

\bibitem{hhh1} S. A. Hartnoll, C. P. Herzog, G. T. Horowitz, \href{https://doi.org/10.1088/1126-6708/2008/12/015}{JHEP12(2008)015}


\bibitem{CPH}C. P. Herzog, \href{https://doi.org/10.1088/1751-8113/42/34/343001}{2009 J. Phys. A: Math. Theor. 42 343001}

\bibitem{SAH}S. A. Hartnoll, \href{https://doi.org/10.1088/0264-9381/26/22/224002}{2009 Class. Quantum Grav. 26 224002}

\bibitem{GR1}S. Gangopadhyay, D. Roychowdhury, \href{https://doi.org/10.1007/JHEP05(2012)002}{J. High Energ. Phys. (2012) 2012: 2.}

\bibitem{GR2}S. Gangopadhyay, D. Roychowdhury, \href{https://doi.org/10.1007/JHEP05(2012)156}{J. High Energ. Phys. (2012) 2012: 156.}


\bibitem{SG1}S. Gangopadhyay, \href{https://doi.org/10.1016/j.physletb.2013.06.027}{Physics Letters B
Volume 724, Issues 1–3, 9 July 2013, Pages 176-181}



\bibitem{GG1}D. Ghorai, S. Gangopadhyay, \href{https://doi.org/10.1140/epjc/s10052-016-4005-0}{Eur. Phys. J. C (2016) 76: 146.}



\bibitem{wald}R. M. Wald \href{https://doi.org/10.1103/PhysRevD.10.1680}{Phys. Rev. D 10, 1680}


\bibitem{KM}S. S. Komissarov, J.C. McKinney, \href{https://doi.org/10.1111/j.1745-3933.2007.00301.x}{MNRAS: Letters, Volume 377, Issue 1, 1 May 2007, Pages L49-L53}

\bibitem{js} J. Sonner,  \href{https://doi.org/10.1103/PhysRevD.80.084031}{Phys. Rev. D 80, 084031}


\bibitem{la} K. Lin, E. Abdalla,  \href{https://doi.org/10.1140/epjc/s10052-014-3144-4}{E. Eur. Phys. J. C (2014) 74: 3144.}


\bibitem{GKS}R. Gregory, S. Kanno, J. Soda, \href{https://doi.org/10.1088/1126-6708/2009/10/010}{JHEP10(2009)010}

\bibitem{as} A. Sheykhi, \href{https://doi.org/10.1103/PhysRevD.78.064055}{Phys. Rev. D 78, 064055}



\bibitem{sg}S. Gangopadhyay,  D. Roychowdhury,  \href{https://doi.org/10.1007/JHEP08(2012)104}{J. High Energ. Phys. (2012) 2012:104.}

\bibitem{BF1}P. Breitenlohner, D. Z. Freedman, \href{https://doi.org/10.1016/0003-4916(82)90116-6}{Ann. Phys. (N.Y.) 144, 249 (1982)}

\bibitem{BF2}P. Breitenlohner, D. Z. Freedman, \href{https://doi.org/10.1016/0370-2693(82)90643-8}{ Phys. Lett. 115B, 197 (1982).}



\bibitem{st}G.~Siopsis, J.~Therrien, \href{https://doi.org/10.1007/JHEP05(2010)013}{J. High Energ. Phys. (2010) 2010: 13.}

\bibitem{rb}R. Banerjee, S. Gangopadhyay, D. Roychowdhury, A. Lala,
\href{https://doi.org/10.1103/PhysRevD.87.104001}{Phys. Rev. D 87, 104001}.

\end{thebibliography}
\end{document}